\documentclass[%
 reprint,
 amsmath,amssymb,
 aps,prc
]{revtex4-2}

\usepackage{graphicx}
\usepackage{dcolumn}
\usepackage{bm}
\usepackage{comment}
\usepackage{mathrsfs}
\usepackage{siunitx}
\usepackage{physics}
\usepackage[pdftex, colorlinks]{hyperref} 
\usepackage{multirow}

\newcommand{\He}{^8\mathrm{He}}
\newcommand{\C}{^{12}\mathrm{C}}
\newcommand{\Be}{^{10}\mathrm{Be}}

\begin{document}
\preprint{APS/123-QED}

\title{$\alpha+{}^2n+{}^2n$ 3-body cluster structures and dineutron breaking in $\He$}

\author{Kosei Nakagawa}
\affiliation{Department of Physics, Kyoto University, Kyoto, 606-8502, Japan}
\author{Yoshiko Kanada-En'yo}
\affiliation{Department of Physics, Kyoto University, Kyoto, 606-8502, Japan}

\date{\today}

\begin{abstract}
\begin{description}
\item[Background]
The $0_2^+$ state of $\He$ has been discovered by a recent experiment, which suggested a developed cluster structure of spatially correlated neutron pairs, called "dineutrons" (${}^2n$).
\item[Purpose]
We aim to investigate the structure of $\He(0^+_1)$ and 
$\He(0^+_2)$ to clarify the monopole excitation mode in $\He$ system while focusing on the 
$\alpha+{}^2n+{}^2n$ cluster structures and the breaking of ${}^2n$ clusters.
\item[Methods]
We apply a microscopic cluster model with the generator coordinate method for the $\alpha+{}^2n+{}^2n$ cluster and $^6{\rm He}+{}^2n$ cluster dynamics.  The
$p_{3/2}$-closure component, which is induced by the spin-orbit force,
is also incorporated.
\item[Results]
The present calculation reasonably reproduces the experimental data
of the properties of $\He$, such as $2n$ and $4n$ separation energies, energy spectra, and radii.
The spatially developed cluster structure of the $0_2^+$ state is obtained.
The $0_1^+$ and $0_2^+$ states have dominant components of 
$\alpha+{}^2n+{}^2n$ 3-body cluster, but contain 
significant ${}^2n$ breaking components which contribute 
to the energy gain and size shrinking of the $\He$ system. 
The monopole excitation in $\He$ is regarded as a 
radial excitation, which is similar to that in $\C$.
\item[Conclusions]
The $\alpha+{}^2n+{}^2n$ cluster structures play a dominant role in $\He(0_{1,2}^+)$, and the mixing of the dineutron breaking contributes to significant structural effects.
\end{description}
\end{abstract}

\maketitle

\section{\label{sec:intro}introduction}
  In nuclear systems, two neutrons have spatial correlation due to the
attractive nuclear interaction, even though the attraction is not enough to bind two neutrons in a vacuum. 
In neutron matter, the neutron pair correlation is strong at dilute densities, as the pair size is smaller than the mean distance of two neutrons~\cite{matsuo2006spatial}, and this phenomenon corresponds to the crossover region between the Bardeen-Cooper-Schrieffer (BCS) and the Bose-Einstein condensation (BEC) state.
Also in finite nuclear systems, the neutron pair correlation is 
enhanced at the nuclear surface~\cite{catara1984relation, pillet2007generic}. These phenomena of strong spatial correlations of 
two neutrons are called "dineutron" formation or correlation.

The dineutron plays a further important role in neutron-rich nuclei
having loosely bound valence neutrons.  
For instance, in two-neutron halo nuclei such as $^6$He
and $^{11}$Li, two valence neutrons spread widely outside a core nucleus and have the spatial correlation forming the dineutron~\cite{esbensen1992soft,vaagen1999dineutron}.
In the case of  $^{11}$Li,
the dineutron correlation contributes to the strong low-energy $E1$ transition~\cite{esbensen1992soft,hagino2005pairing,nakamura2006observation}.
Dineutron behaviors and multi-dineutron phenomena 
also attract attention as discussed  
in tetraneutrons~\cite{duer2022observation,hiyama2016possibility,lazauskas2023energy}
 and neutron-skin nuclei. 

$\He$ has four valence neutrons around an $\alpha$ core
and can be a good example of a multi-dineutron system. 
In works for the ground state of $\He$~\cite{hagino2008,yamaguchi2023}, 
the formation of two dineutrons was suggested, even though the shell-model (SM) $p_{3/2}$-closure configuration still contributes to the ground state.
In other words, it is essential
to take into account both the dineutron formation and its breaking due to spin-orbit interaction, in particular, for the ground state of $\He$.

The dineutron (${}^2n$) formation may play further important roles in excited states of $\He$.  
Theoretical studies with the
anti-symmetrized molecular dynamics (AMD) model~\cite{kanada2007n2} and that with the $^2n$ condensation model~\cite{kobayashi2013} predicted 
the second $0^+$ state of $\He$ and suggested the spatially developed $\alpha + {}^2n + {}^2n$ 3-body cluster structure.
This $\He(0_2^+)$ was regarded
as a 3-body cluster gas state in analogy to the $3\alpha$ cluster gas of $^{12}$C$(0_2^+)$.
$\He(0_2^+)$ was also studied 
with $\alpha + 4n$ 5-body model~\cite{myo2010fivebody}, but the dineutron formation was not obtained. 
Recently, an experimental search for 
new states of $\He$ was performed by inelastic scattering 
in inverse kinematics and successfully 
observed $\He(0_2^+)$ for the first time~\cite{yang2023}. 
Moreover, the observed 
matrix element of the isoscalar monopole (IS0) transition between the  $0_1^+$ and $0_2^+$ states was large enough to support the developed cluster structure in the $0_2^+$ state, because the strong IS0 transition is a signal for cluster states as pointed out by Yamada $et$ $al.$~\cite{ISMyamada}.

This paper aims to investigate structures of
the $0_1^+$ and $0_2^+$ states of $\He$ and reveal 
$\alpha+{}^2n+{}^2n$ cluster dynamics
while taking into account the breaking of dineutron clusters.
For this aim, we apply a microscopic model based on a cluster model combined with cluster breaking configurations.
In this model, the 3-body cluster dynamics is described in detail 
within the cluster generator coordinate method (GCM)~\cite{GCM1, GCM2}, and the cluster breaking configurations induced by spin-orbit interactions are incorporated by using the anti-symmetrized quasicluster model (AQCM) proposed by Itagaki $et$ $al.$~\cite{aqcmganso,aqcm}.
We also calculate $\C$ with a similar method as done by Suhara $et$ $al. $ in Ref.~\cite{suhara2015} to discuss 
similarities of 3-body cluster structures in $\He$ to those in $\C$.

This paper is organized as follows.
Section \ref{sec:form} explains the formulation, and Sec.~\ref{sec:res} presents
the calculated results of $\He$, $\C$, and $\Be$. Structure properties
are shown, and the 3-body cluster dynamics and cluster breaking effects are discussed in Sec.~\ref{sec:disc}.
Finally, a summary is given in Sec.~\ref{sec:sum}. 

\section{\label{sec:form}formulation}
  \subsection{\label{form.gene}general formulation}
To describe relative motions between the $\alpha+{}^2n+{}^2n$ clusters in $\He$, we adopt a microscopic cluster model~\cite{BBwf} combined with the generator coordinate method~(GCM)~\cite{GCM1,GCM2} for the $\alpha + {^2n}+{^2n}$ cluster.
The wave function in our model is given by a linear combination of Brink-Bloch~(BB) wave functions~\cite{BBwf}.
In addition to the $\alpha + {^2n}+{^2n}$ cluster configurations, we mix dineutron breaking configurations within our microscopic model formalism because dineutrons are fragile and can be easily dissolved at the surface of the $\alpha$ cluster due to the spin-orbit interaction.
We consider one-dineutron and two-dineutron breaking configurations. 
The former is expressed by the 2-body BB wave functions of the ${}^6\mathrm{ He}+{}^2n$ clusters where the ${}^6\mathrm{ He}$-cluster has $(0s)^4(0p_{3/2})^2$ configurations (simply denoted as "$^6\mathrm{He}$" in this paper).
The latter is the $(0s_{1/2})^4(0p_{3/2})^4$ configuration with the neutron $0p_{3/2}$-closure, which is the lowest configuration 
of the $jj$-coupling shell model.
For numerical simplicity, we use the same width parameter for dineutron and $\alpha$ clusters, and the 
harmonic oscillator of shell-model configurations. In the present framework, the center of mass motion is exactly removed. 

For $\C$, we adopt the microscopic $3\alpha$ cluster model + GCM with the mixing of  $p_{3/2}$-closure configuration to take into account the cluster breaking effect induced by the spin-orbit interaction.  
For $\Be$, we apply the microscopic $2\alpha+{^2n}$ cluster model + GCM and the incorporate the dineutron 
breaking component by mixing  $\mathrm{^6He}+\alpha$ 2-body cluster configurations.

\subsection{\label{form.wf}model wave functions}
We consider a three cluster system composed of clusters $C_i$~$(i=1,2,3)$ with the mass numbers $A_i$, 
and express the 3-body cluster wave function of $C_i$ centering at $\bm{R}_i$ as,
\begin{align}
\label{3body}
   &\ket{C_1,C_2,C_3;\bm{R}_1,\bm{R}_2,\bm{R}_3}  \notag  \\
  & \quad \quad \quad = {\cal A}  \Bigl[ \ket{C_1;\bm{R}_1}  \ket{C_2;\bm{R}_2} \ket{C_3;\bm{R}_3} \Bigr], 
\end{align}
where $\ket{C_i;\bm{R}_i}$ is the $i$th cluster wave function and ${\cal A}$ is the antisymmetrizer.

For 3-body cluster configurations of $\He$, $C_i$s correspond to $\alpha$ and ${}^2n$ clusters,
and are given by the $(0s)^4$ and $(0s)^2$ wave functions as,
\begin{gather}
  \ket{\alpha;\bm{R}_i} = {\cal A}   \left[ \prod_{j=1}^4 e^{-\nu(\bm{r}_j -\bm{R}_i)^2} 
  \ket{p \uparrow}_1 \ket{p \downarrow}_2 \ket{n \uparrow}_3 \ket{n \downarrow}_4 \right],
\end{gather}
\begin{equation}
  \ket{{}^2n;\bm{R}_i} ={\cal A}  \left[ \prod_{j=1}^2 e^{-\nu(\bm{r}_j -\bm{R}_i)^2} 
   \ket{n \uparrow}_1 \ket{n \downarrow}_2 \right],
\end{equation}
where $p\uparrow$($\downarrow$) and $n\uparrow$($\downarrow$)  denote the spin-up (down) proton and neutron.

Similarly, 2-body cluster wave functions are also expressed as,
\begin{equation}\label{2body}
  \ket{C_1,C_2;\bm{R}_1,\bm{R}_2}  = {\cal A}  \left[ \ket{C_1;\bm{R}_1} \ket{C_2;\bm{R}_2} \right].
\end{equation}

The 3-body and 2-body cluster configurations are projected onto the parity and total angular momentum $J^\pi=0^+$ states.
Furthermore, to exactly remove the center of mass motion of the total $A$-nucleon system of $k$-body cluster wave functions ($A= \sum_{i=1}^{k} A_i$),
we set the condition 
\begin{equation}
  \bm{R}_G \equiv \frac{1}{A} \sum_{i=1}^{k} A_i \bm{R}_i=0.
\end{equation}

In order to construct the $p_{3/2}$-closure configuration and the ${^6\rm He}$-cluster wave functions used in the 2-body cluster configurations,
we use the expression of the anti-symmetrized quasicluster model~\cite{aqcmganso, aqcm} proposed by Itagaki $et$ $al$. 
In this model, $p_{3/2}$ orbits are expressed by infinitesimally shifted Gaussians.

Practically, all of $A$-nucleon wave functions used in the present calculation are written in Gaussian form. They are expressed by 
Slater determinants of single-particle Gaussian wave functions within the AMD framework~\cite{Kanada1995Clustering, Kanada2012Antisymmetrized, AMDN6} as 
\begin{eqnarray}
  \Phi_{\rm AMD} = {\rm det}\left[\psi_1\psi_2\cdots\psi_A\right],
\end{eqnarray}
where $\psi_i$ represents the $i$th single-particle wave function written by a product of spatial, 
spin, and isospin functions as follows:
\begin{eqnarray}
\psi_i\ &=& \phi(\bm{Z}_i)\chi(\bm{\xi}_i)\tau_i, \\
\phi(\bm{Z}_i) &=& \left(\frac{2\nu}{\pi}\right)^{\frac{3}{4}}\exp\left[ -\nu\left(\bm{r} - \frac{\bm{Z}_i}{\sqrt{\nu}}\right)^2 \right],\\
\chi(\bm{\xi}_i)&=&\xi_{i\uparrow}|\uparrow\rangle + \xi_{i\downarrow}|\downarrow\rangle, \\
\tau_i &=&p\ \textrm{or}\ n.
\end{eqnarray}
Here, $\bm{Z}_i$ and $\bm{\xi}_i$ are complex parameters for Gaussian centers and spin directions, respectively. 
By setting specific values for $\bm{Z}_i$ and $\bm{\xi}_i$ parameters, we can express 
the 3-body, 2-body, and $p_{3/2}$-closure configurations. 
For more detailed expression of the $(0s)^4(0p_{3/2})^2$ configurations of the $^6\rm{He}$-cluster in the 2-body cluster configurations 
and the $p_{3/2}$-closure configuration in the form of AMD with AQCM, the reader is referred to Refs.~\cite {aqcm, aqcmganso, mono10Be9Li}.

\subsection{\label{form.GCM}3-body and 2-body cluster GCM + configuration mixing}
For 3-body cluster configurations, we use hyperspherical coordinates~\cite{descouvemont2019,suno2015,hsc2004} for 
the positions $\bm{R}_i$ of three clusters $C_i$($i=1,2,3$) in Eq.~\eqref{3body}. 
The scaled Jacobi coordinates $\bm{X}$ and $\bm{Y}$ are defined as 
\begin{align}
 & \bm{X} = \sqrt{\frac{A_1A_{2}}{A_{12}}} \left(\bm{R}_1-\bm{R}_2\right) , \quad \\
 & \bm{Y} = \sqrt{\frac{A_{12}A_3}{A}} \left(\frac{A_1\bm{R}_1+A_2\bm{R}_2}{A_{12}}-\bm{R}_3\right),
\end{align}
with $A_{12}=A_1+A_2$. Then, the intrinsic configurations before the $J^\pi=0^+$ projection are specified by 3 parameters of 
$X=|\bm{X}|$, $Y=|\bm{Y}|$, and the angle $\theta$ between $\bm{X}$ and $\bm{Y}$.
We introduce hyperradial $\rho$ and hyperangle $\gamma$ instead of $X$ and $Y$ as,
\begin{align}
  \rho = \sqrt{\bm{X}^2 + \bm{Y}^2}, \quad \gamma = \frac{Y}{X}.
\end{align}
Note that the hyperradial is defined independently of the choice of cluster labels $i=1,2,3$, and hence it is a useful measure of 
the system size. 
By using $(\rho, \gamma , \theta)$, we rewrite the 3-body cluster configuration equivalent to that of Eq.~\eqref{3body} as, 
\begin{equation}
  \ket{C_1,C_2,C_3;\bm{R}_1,\bm{R}_2,\bm{R}_3} \to \ket{C_1,C_2,C_3;\rho,\gamma,\theta}.
\end{equation}

For 2-body cluster configurations, we introduce the distance parameter $a\equiv|\bm{R}_1-\bm{R}_2|$
to specify the intrinsic configurations and rewrite Eq.~\eqref{2body} as 
\begin{equation}
  \ket{C_1,C_2;\bm{R}_1,\bm{R}_2}  \to  \ket{C_1,C_2;a}.
\end{equation}

In the GCM calculations, we superpose the 3-body cluster configurations
using the generator coordinates $\rho$, $\gamma$, and $\theta$, and 
the 2-body cluster configurations with the generator coordinate $a$. 
Then, the total wave function of $\He$ is given by superposition of $J^{\pi}=0^+$ projected configurations of the $\alpha+{}^2n+{}^2n$, and the
${}^6{\rm He}+{}^2n$ cluster states with mixing of the shell-model~(SM) configuration $\ket{\Psi^{SM}}$
of the $(0s_{1/2})^4(0p_{3/2})^4$ state. 
Thus, the $\He$ wave function of the $\nu$th state is written as 
\begin{equation}
  \label{wf.int}
\begin{split}
&\ket{\He(0_\nu ^+)} \\
&= \int d\rho \, d\gamma \, d\theta \, c^{3B}_{\nu}(\rho,\gamma,\theta) \hat{P}^{0+}\ket{\alpha+ {^{2}n}+ {^{2}n};\rho,\gamma,\theta} \\
&+ \int da \, c_{\nu}^{2B}(a) \hat{P}^{0+}\ket{{^{6} \rm He}+{^{2}n};a} + c_{\nu}^{SM} \ket{\Psi^{SM}},
\end{split}
\end{equation}
where $\hat{P}^{J\pi}$ is the $J^\pi$ projection operator, and coefficients $c_{\nu}^{3B}$, $c_{\nu}^{2B}$, and $c_{\nu}^{SM}$ are determined by variational principle.
Here, three clusters are chosen as $C_1=C_2={}^2n$ and $C_3=\alpha$ to save the range of $\theta$.

In the practical calculation, 
the integrations for the generator coordinates are performed by 
summation of discretized points, and Eq.~\eqref{wf.int} is rewritten in a discretized form as 
\begin{equation}
\label{wf.disc}
  \begin{split}
    \ket{\He (0_\nu^+)} &= \sum_{i} c_{\nu}^{3B}(i) \ket{\Psi_i^{3B}} + \sum_{i} c_{\nu}^{2B}(i) \ket{\Psi_i^{2B}} \\
    &+ c_{\nu}^{SM} \ket{\Psi^{SM}},
  \end{split}
\end{equation}
using the $0^+$-projected configurations of the discretized bases,
\begin{equation}
\label{3Bbase.dis}
    \ket{\Psi_i^{3B}} \equiv \hat{P}^{0+}\ket{\alpha+ {}^2n+ {}^2n;\rho=\rho_i,\gamma=\gamma_i,\theta=\theta_i},
\end{equation}
\begin{equation}
    \label{2Bbase.dis}
    \ket{\Psi_i^{2B}} \equiv \hat{P}^{0+}\ket{^6\mathrm{He}+{}^2n;a=a_i}.
\end{equation}
Here, the coefficients $c_{\nu}^{3B}(i)$,  $c_{\nu}^{2B}(i)$, and  $c_{\nu}^{SM}$ are determined by 
solving the generalized eigenvalue problem of the Norm and Hamiltonian matrices, which is derived from the variational principle.

For configurations of $\ket{\Psi^{3B}_i}$, 485 sets of 
$(\rho_i,\gamma_i,\theta_i)$ with $\rho_i =1, 2, \cdots, \SI{12}{fm}$, 
$\gamma_i = 0, \pi/2\rho, \cdots $  $(\gamma \le \pi/2$), 
$\theta_i = 0, \pi/Y, \cdots, \pi/2$ are adopted. 
For $\ket{\Psi_i^{2B}}$, $a_i=1,2,\cdots, \SI{12}{fm}$ are used.

Similarly, 
the total wave function of ${}^{12}{\rm C}$ is obtained by superposing $3\alpha$ configurations and the SM $p_{3/2}$-closure configuration $\ket{(0p_{3/2})^8}$, and that of $\Be$ is obtained by 
superposing $2\alpha+{}^2n$ and ${}^6{\rm He}+\alpha$ configurations as 
\begin{equation}
\begin{split}
\label{wf.12C}
\ket{\C (0_{\nu}^+)} 
&=\int d\rho \, d\gamma \, d\theta \, c_{\nu}^{3B}(\rho,\gamma,\theta) \hat{P}^{0+}\ket{3\alpha;\rho,\gamma,\theta} \\
&+ c_{\nu}^{SM} \hat{P}^{0+}\ket{(0p_{3/2})^8}, 
\end{split}
\end{equation}
\begin{equation}
  \begin{split}
  \label{wf.10Be}
&\ket{\Be(0_\nu^+)}  \\
&=\int d\rho \, d\gamma \, d\theta \, c_{\nu}^{3B}(\rho,\gamma,\theta) \hat{P}^{0+}\ket{2\alpha+{^{2}n};\rho,\gamma,\theta} \\
&+ \int da c_{\nu}^{2B}(a) \hat{P}^{0+}\ket{\mathrm{^{6}He}+\alpha;a}.    
  \end{split}
\end{equation}
For the $2\alpha+{}^2n$ configurations of $\Be$, we choose $C_1=C_2={}^2n$, $C_3=\alpha$. 
The integrations of the generator coordinates are performed using the same discretized points as done in $\He$, and the coefficients are determined respectively for $\C$ and $\Be$ as well.

\subsection{\label{form.hamil}Hamiltonian}
  The Hamiltonian is composed of the kinetic energy $t_i$ of the $i$th particle, the effective nuclear forces including the central force $V_{ij}^{c}$ and the spin-orbit force $V_{ij}^{ls}$, and the Coulomb force $V_{ij}^{Coul}$ between particles $i$ and $j$:
\begin{equation}
  H = \sum_{i=1}^A t_i - t_G + \sum_{i<j} V_{ij}^{c} + V_{ij}^{ls} + V_{ij}^{Coul},
\end{equation}
where $t_G$ is the kinetic energy of the center of mass.
As for the central force, we use Volkov No.2~\cite{VOLKOV196533}, and choose the parameters $M=0.58$ and $B=H=0$ for $\He$ to reproduce overall properties of $\mathrm{He}$ isotopes~\cite{aoyama2006}.
For $\C$ and $\Be$, $M=0.60$ and $B=H=0$ are adopted.
\begin{table*}[!t]
  \centering
  \caption{The binding energies (B.E.), $2n$ and $4n$ separation energies $(S_{2n,4n})$,
   and $0_2^+$ excitation energies ($E_x(0_2^+)$) of $\He$.
  The present results are listed together with the experimental data
  ~\cite{energy8he, yang2023},  theoretical values of AMD model~\cite{kanada2007n2}, and
    those of the ${}^2n$ condensation model~\cite{kobayashi2013}. Units are in MeV. 
}
  \label{tab:he8E}
  \begin{tabular}{ccccc}
  \hline \hline
                                               & B.E. & $S_{2n}$&$S_{4n}$  & $E_x(0_2^+)$  \\
   \hline 
    This work                                  & 29.44  &  1.47  &  1.83  & 5.20      \\
    Expt.                       & 31.60\cite{energy8he} &  2.14\cite{energy8he}&  3.12  & 6.66(6)\cite{yang2023}      \\
    AMD              ~\cite{kanada2007n2}  & 32.1   &  3.0   & 4.3   & 10.3      \\
    ${}^2n$ cond.~\cite{kobayashi2013} & 30.98   &  2.51  &  3.6  & 7.9  \\
    \hline \hline
  \end{tabular}
\end{table*}
\begin{table*}
  \centering
  \caption{
   R.m.s. matter radii $r_m$, point proton (neutron) r.m.s. radii $r_p$ ($r_n$) 
  of $\He(0_1^+)$ and $\He(0_2^+)$, and IS0 matrix elements $M$(IS0) for the $0_1^+$-$0_2^+$ transition.
  The values of the present calculation, the experimental data~\cite{mat.rad.8he, nrad.8he},
   and the other theoretical calculations~\cite{kanada2007n2,kobayashi2013} are listed. 
The experimental point proton radii $r_p$ are evaluated using the relation
 $r_p^2 = r_c^2 - R_p^2 -\frac{3}{4m_p^2}-\frac{N}{Z}R_n^2$ with the charge radius
  $r_c$ of $\He$~\cite{ch.rad.8he} and the charge radius of a single proton (neutron) $R_{p(n)}$~\cite{PDG2024}.}
  \label{tab:he8rad}
  \begin{ruledtabular}
  \begin{tabular}{cccccccc}
                           &\multicolumn{3}{c}{$0_1^+$} &\multicolumn{3}{c}{$0_2^+$} & \multirow{2}{*}{$M$(IS0) (\si{fm^2})}\\
        \cline{2-7}
             &  $r_m$ $ (\si{fm})$   & $r_p$ $ (\si{fm})$  &\multicolumn{1}{c}{$r_n$ $ (\si{fm})$}&  $r_m$ $ (\si{fm})$ & $r_p$ $ (\si{fm})$  &\multicolumn{1}{c}{$r_n$ $ (\si{fm})$}  &  \\
   \hline 
      This work                                 &  2.40  &   1.89  &  2.55  &  3.91  &  2.64  &  4.25  & 15.5 \\
       Expt.               &2.45(7)\cite{mat.rad.8he} &1.82(3)&  2.69(4)\cite{nrad.8he}&&             &&$11^{+1.8}_{-2.3}$\\
     AMD              ~\cite{kanada2007n2}  &  2.24  &  1.76   &  2.36  &  2.73  &  1.97  &  2.94  &7.3\\
    ${}^2n$ cond.~\cite{kobayashi2013} &  2.49  &  1.93   &  2.65  &  4.67  &  2.21  & 5.24   &10.3\\
  \end{tabular}
  \end{ruledtabular}
\end{table*}
As $V_{ij}^{ls}$, we use the spin-orbit part of the G3RS interaction~\cite{G3RS, G3RStamagaki} 
\begin{equation}
  V_{ij}^{ls} = \sum_{k=1}^2 u_k e^{-(r_{ij} / \xi_k)^2} P(^3 O) \bm{L}_{ij} \cdot \bm{S}_{ij},
\end{equation}
where $P(^3 O)$ is a projection operator onto the triplet-odd state.
We choose $u_1=-u_2= \SI{1600}{MeV}$ for all the nuclei.
This set of the interaction for $\C$ and $\Be$ is the same as that used in previous researches~\cite{suhara2015, mono10Be9Li, AMDN6}.

\section{\label{sec:res}result}
  \subsection{Structure properties of \texorpdfstring{$\He$, $\C$ }  aand \texorpdfstring{$\Be$}{He, C, Be}}
The calculated energies for $\He(0^+_{1,2})$ are shown in Table~\ref{tab:he8E}.
The binding energies (B.E.), two-(four-)neutron separation energies $S_{2n(4n)}$
of $\He(0^+_1)$, and the excitation energy $E_x(0_2^+)$ of
$\He(0^+_2)$ are listed together with those of the experimental data~\cite{energy8he, yang2023} and theoretical values of AMD~\cite{kanada2007n2} and ${}^2n$ condensation~\cite{kobayashi2013} model calculations. 
Our calculation reasonably reproduces the binding energy systematics of He isotopes with 
the present choice of interaction parameters.
For the $0_2^+$ state, the relative energy from the $\alpha+4n$ threshold
 is obtained as $\SI{3.37}{MeV}$ in the present result, 
 which is in good agreement with the experimental value $\SI{3.54}{MeV}$ and also 
 consistent with the theoretical value $\SI{4.3}{MeV}$ of the ${}^2n$ condensation model~\cite{kobayashi2013}.

\begin{table}[t]
  \centering
  \caption{Energies, r.m.s. radii, and IS0 transition matrix elements of $\C$ obtained with the present $3\alpha+(0p_{3/2})^8$ calculation. 
The experimental data~\cite{kelley2017energy,ozawa2001nuclear,chernykh2010pair} and 
the other theoretical values of the $\alpha$ condensation model~\cite{THSR, funaki2003analysis} and AMD model~\cite{Kanada2007Structure} are also shown. 
The experimental value of the IS transition matrix element 
$M({\rm IS0})$
is deduced from the $E0$ transition matrix element $M(E0)$ 
assuming $M({\rm IS0})=2M(E0)$.}
  \label{tab:c12res}
  \begin{ruledtabular}
\begin{tabular}{cccccc}
                                          &  B.E.  &  $E_x(0_2^+)$  &  $r_m(0_1^+)$  &  $r_m(0_2^+)$  &  $M$(IS0) \\
                                          &  (MeV)       &  (MeV)       &  (fm)          &  (fm)          &  ($\si{fm^2}$)\\
               \hline  
     This work                            &  90.02      &  8.04      &  2.38          &  3.18          &  16.14      \\
     Expt.                                &  92.16      &  7.65\cite{kelley2017energy}      &  2.35(2)\cite{ozawa2001nuclear}       &    &  10.8\cite{chernykh2010pair}   \\
     $\alpha$ cond.~\cite{funaki2003analysis}&  89.52   &  7.7       &  2.40          &  3.47          &  13.4    \\
     AMD~\cite{Kanada2007Structure}       &  88.0       &  8.1       &  2.53          &  3.27          &  13.4 \\
  \end{tabular}
  \end{ruledtabular}
\end{table}

The calculated results for radii and the isoscalar monopole (IS0) transition of $\He$ are shown in Table~\ref{tab:he8rad}. 
The values of the root mean squared (r.m.s.) matter radius $r_m$, point proton and neutron radii $r_{p,n}$ of $\He(0_1^+)$ and $\He(0_2^+)$ are listed compared with the experimental data~\cite{mat.rad.8he, ch.rad.8he, nrad.8he}, and the other theoretical calculations~\cite{kanada2007n2, kobayashi2013}.
The values of the IS0 transition matrix element $M$(IS0) between the $0_1^+$ and $0_2^+$ states are also listed in the table.
Here, the IS0 transition operator is defined as, 
\begin{align}
  \sum_{i=1}^A (\bm{r}_i-\bm{r}_g)^2,
\end{align}
where $\bm{r}_g=\frac{1}{A} \sum_{i=1}^A \bm{r}_i$ is the center-of-mass coordinate.

The present results of $r_{m}$, $r_{p}$, and $r_n$ for the $0_1^+$ state are in good agreement with the experimental values, and also consistent with the other calculations.
For the $0_2^+$ state, we obtain larger radii than those of the $0_1^+$ state because of 
the spatially spreading $\alpha+{}^2n+{}^2n$ cluster structure of $\He(0_2^+)$.
In particular, $r_n(0_2^+)$ 
is significantly large due to the developed dineutron cluster structure.

\begin{table*}
  \centering
  \caption{
Energies, r.m.s. radii, and IS0 transition matrix element of $\Be$
obtained with the present calculation using the $2\alpha+{}^2n$ and $^6{\rm He}+ \alpha$ configurations.
The experimental data~\cite{energy8he,tanihata1985measurements,nortershauser2009nuclear} and the other theoretical values of the MO model~\cite{10BeMO,10BeMO2} and the $^6\mathrm{He}+\alpha$ model~\cite{mono10Be9Li, 9Li2body} are also listed.
}
  \label{tab:Energy10Be}
  \begin{ruledtabular}
  \begin{tabular}{ccccccc}
                                                        &  B.E.  &  $E_x(0_2^+)$  &  $r_m(0_1^+)$  &  $r_p(0_1^+)$  &  $r_m(0_2^+)$  &  $M$(IS0)\\
                                                        &  (MeV)       &  (MeV)       &  (fm)          &  (fm)          &  (fm)          &  ($\si{fm^2}$)\\
               \hline  
    This work                                           &  60.59      &  5.88      &  2.52          &  2.45          &  3.34          &  15.18   \\
    Expt.                                               &  64.98      &  6.18\cite{energy8he}       &  2.39(2)\cite{tanihata1985measurements}      &  2.21(2)\cite{nortershauser2009nuclear}       &                &      \\
    MO\cite{10BeMO,10BeMO2}                              &  61.4       &  8.1       &                &  2.51          &                &  \\
    $^6\mathrm{He}+\alpha$~\cite{mono10Be9Li, 9Li2body}  &  56.9       &  8.3      &  2.34          &  2.31          &                &  10.0  \\
  \end{tabular}
    \end{ruledtabular}
\end{table*}

\begin{table*}
  \centering
  \caption{Proportions of the $\alpha+{}^2n+{}^2n$ 3-body cluster, the SM $(p_{3/2})^4$, and the $\mathrm{^6He}+{}^2n$ 2-body cluster components in $\He(0_1^+)$ and $\He(0_2^+)$, 
those of the $3\alpha$ 3-body cluster and the SM components in $\C(0_1^+)$ and $\C(0_2^+)$, 
and those of the $2\alpha+{}^2n$ 3-body cluster and $\mathrm{^6He}+{}^2n$ 2-body cluster components in  $\Be(0_1^+)$ and $\Be(0_2^+)$.}
  \label{tab:123percentage}
  \begin{ruledtabular}
  \begin{tabular}{ccccccc}
                      &$\He(0_1^+)$&$\He(0_2^+)$&$\C(0_1^+)$&$\C(0_2^+)$&$\Be(0_1^+)$&$\Be(0_2^+)$ \\
                 \hline 
       3-body         &   0.841    &    0.926   &  0.858    &    0.864  &  0.887     &  0.924    \\
     SM               &   0.487    &    0.086   &  0.293    &    0.170   \\
       2-body         &   0.883    &    0.583   &           &           &  0.834      &  0.628 \\
  \end{tabular}
  \end{ruledtabular}
\end{table*}

For the IS0 $0_1^+ - 0_2^+$ transition,
 the calculated matrix element is as large as $M$(IS0)=$\SI{15.5}{fm^2}$, which is mainly a neutron contribution due to the developed dineutron clusters. 

Structures of $\C(0_1^+)$ and $\C(0_2^+)$ are calculated in the present framework using the 3$\alpha$ and $(0p_{3/2})^8$ configurations given in Eq.~\eqref{wf.12C}. 
The results of the energies, radii, and the IS0 transition matrix element are shown in Table \ref{tab:c12res} compared with the experimental values~\cite{kelley2017energy,ozawa2001nuclear,chernykh2010pair} and other theoretical values of 
the $\alpha$ condensation model~\cite{THSR,funaki2003analysis} and the AMD model~\cite{Kanada2007Structure}.
As shown in Table \ref{tab:c12res}, overall properties such as B.E.,
 $r_m(0_1^+)$, and $E_x(0^+_2)$ of the present result are in good agreement with experimental data and other theoretical calculations.
In the present result, the radius $r_m$ of $\C(0_2^+)$ is significantly large because of the spatially spreading $3\alpha$ cluster structure.
This is a similar feature to $\He$, but comparing 
$\C$ with $\He$, $r_m$ of $\C(0_2^+)$ is smaller than that of $\He(0_2^+)$ as three clusters are more deeply bound in $\C(0_2^+)$ than in $\He(0_2^+)$. 
For the monopole transition matrix element $M$(IS0),
 the present calculation somewhat overestimates the experimental value. 
The reason for this overestimation might be that
 the present model space contains
  only $3\alpha$ and $(0p_{3/2})^8$ configurations
   but not other cluster breaking configurations.

We also calculate structures of $\Be(0_1^+)$ and $\Be(0_2^+)$ using
 the $2\alpha+{}^2n$ and $^6{\rm He}+\alpha$ configurations given in Eq.~\eqref{wf.10Be}. 
The results for the energies, r.m.s. radii, and IS0 transition matrix element are shown in Table \ref{tab:Energy10Be} together with the experimental values and the other theoretical values of the molecular orbital (MO) model~\cite{10BeMO,10BeMO2} and the $^6\mathrm{He}+\alpha$ model~\cite{mono10Be9Li, 9Li2body} calculations. 
The present results are consistent with the results of the other theoretical works.

\subsection{The 3-body cluster components and the cluster breaking in $\He$, $\C$, and $\Be$}
As explained in Sec.~\ref{sec:form}, the total wave function of $\He$ 
is composed of the $\alpha+{}^2n+{}^2n$ configurations, the $\mathrm{^6He}+{}^2n$ configurations, and the SM configuration. 
The latter two configurations describe
dineutron cluster breaking caused by the spin-orbit interaction around
the $\alpha$ core;
the $\mathrm{^6He}+{}^2n$ configurations and the SM configuration express one- and two-dineutron breaking, respectively.
To evaluate each contribution of the three components contained in the obtained $\He(0^+_1)$ and $\He(0^+_2)$ wave functions, we define the projection operators onto subspaces as follows.

For the $\alpha+{}^2n+{}^2n$ 3-body component in $\He$, 
we construct a set of orthonormal bases $\{ \tilde \Psi^{3B}_i\}$
from a linear transformation of $\{\Psi^{3B}_i \}$ in Eq.~\eqref{3Bbase.dis} and define the projection 
operator $\hat P^{3B}$ of the $\alpha+{}^2n+{}^2n$ subspace as,
\begin{equation}
  \hat P^{3B} = \sum_i \ket{\tilde \Psi_{i}^{3B}}\bra{\tilde \Psi_{i}^{3B}}.
\end{equation}
For $\{ \tilde \Psi^{3B}_i\}$, we adopt eigenvectors of the norm matrix of $\{\Psi^{3B}_i \}$.
We also define the projection operator $\hat P^{2B}$ of the $\mathrm{^6He}+{}^2n$ 2-body subspace 
\begin{equation}
  \begin{split}
    \hat P^{2B} = \sum_i \ket{\tilde \Psi_i^{2B}}\bra{\tilde \Psi_i^{2B}}
  \end{split}
\end{equation}
by transforming $\{\Psi^{2B}_i\}$ in Eq.~\eqref{2Bbase.dis} to an orthonormal bases set $\{ \tilde \Psi^{2B}_i\}$.
The projection operator $\hat P^{SM}$ onto the SM subspace is gained from the normalized base $\ket{\tilde\Psi^{SM}}$ of $\ket{\Psi^{SM}}$ in Eq.~\eqref{wf.int}, and given as
\begin{equation}
  \hat P^{SM} = \ket{\tilde \Psi^{SM}}\bra{\tilde \Psi^{SM}}.
\end{equation}
We evaluate the proportion of each subspace component by calculating 
the expectation values of the projection operators ${\hat P}^{\textrm{sub}}={\hat P^{3B}},{\hat P^{2B}}, {\hat P^{SM}}$ for $\He(0_1^+)$ and $\He(0_2^+)$ as 
\begin{equation}
 \expval{\hat P^{\textrm{sub}}}{\He(0_k^+)}, \quad k=1,2.
\end{equation}
Note that each configuration subspace is not orthogonal to each other, so the sum of the proportions exceeds 1.

Similarly, we calculate the proportions of the 3$\alpha$ and SM components in $\C$ and 
the $2\alpha+{}^2n$ and $^6{\rm He}+\alpha$ components in $\Be$ by constructing the 
corresponding subspace operators.
The calculated results for each component in $\He(0_{1,2}^+)$, 
$\C(0_{1,2}^+)$, and $\Be(0_{1,2}^+)$ are shown 
in Table \ref{tab:123percentage}.

In $\He(0^+_2)$, $\alpha+{}^2n+{}^2n$ component 
is $93 \%$, and therefore, this state is regarded
as the $\alpha+{}^2n+{}^2n$ cluster state.
The $\He(0^+_1)$ state is dominated by the 
$\alpha+{}^2n+{}^2n$ component as $84 \%$, 
but it also has a large overlap with the SM component as $49 \%$. 
It indicates that the $\He(0^+_1)$ state is a mixture of 
the dineutron cluster and the SM component. 

The dineutron cluster breaking component can be evaluated by 
subtracting the $\alpha+{}^2n+{}^2n$
component from 100\%. It corresponds to 
the residual fraction beyond the $\alpha+{}^2n+{}^2n$ subspace
and indicates a non-3-body cluster component.
The dineutron breaking is calculated as 16\% in $\He(0^+_1)$ and 7\% in $\He(0^+_2)$ from the values $84\%$ and $93\%$ of the $\alpha+{}^2n+{}^2n$ component.
It is indicated that significant dineutron breaking occurs in the ground state
because $\He(0^+_1)$ has a compact structure, 
in which dineutron clusters are dissolved by the spin-orbit interaction from the $\alpha$ core. 
On the other hand, the dineutron breaking component is only 7\% in the 
$0^+_2$ state because the spin-orbit interaction is weaker  
for spatially spreading dineutron clusters.
The percentages are not large, but these dineutron breakings give significant effects on the 
structure properties of $\He(0^+_1)$ and $\He(0^+_2)$
such as energies and radii.
The details of the dineutron breaking effects are discussed later. 

For the $\mathrm{^6He}+{}^2n$ component in $\He$,
the $0_1^+$ state has an approximately 90\% overlap, which mainly 
originates in the large overlap of the $\mathrm{^6He}+{}^2n$ component 
with the $\alpha+{}^2n+{}^2n$ and SM components in the spatially 
compact state. 
The $0^+_2$ states have an approximately 60\% overlap with the 
$\mathrm{^6He}+{}^2n$ cluster component, 
indicating the strong spatial correlation 
between one dineutron cluster and the $\alpha$ cluster.

In $\C(0_1^+)$ and $\C(0_2^+)$, the $3\alpha$ cluster component are as large as 86\% in the both states.
$\C(0_1^+)$ has a significant overlap with the SM 
component, indicating that it is a mixture of 
the dominant $3\alpha$ cluster structure and the SM component.
The $\alpha$ cluster breaking components in 
$\C(0_1^+)$ and $\C(0_2^+)$ are found to be $14\%$, which are calculated from 
the non-3-body cluster component by subtracting the $3\alpha$ component.
Compared with $\He$, 
the cluster breaking component in $\C(0_1^+)$ is comparable
to $\He(0^+_1)$, whereas that in $\C(0_2^+)$  is twice larger than
$\He(0^+_2)$. It means that the larger cluster breaking occurs 
in $\C(0_2^+)$, which has the smaller system size and therefore
the stronger spin-orbit interaction effect 
than the case of 
$\He(0^+_2)$.

\begin{figure}[t]
  \centering
  \includegraphics[width=\linewidth]{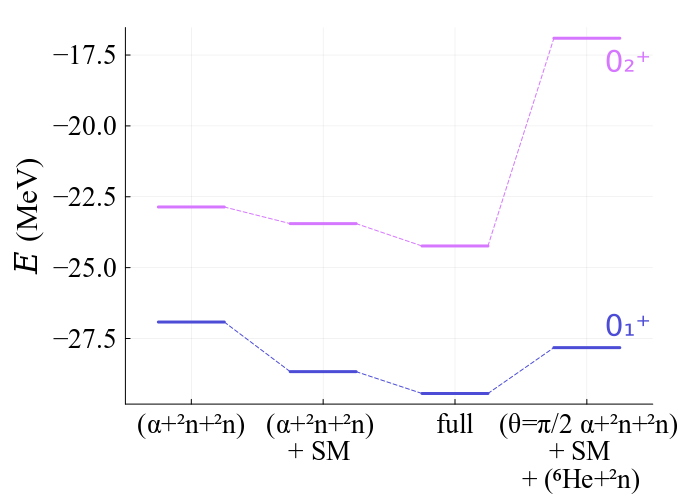}
  \caption{The energy spectra of $\He(0_1^+)$ and $\He(0_2^+)$ calculated with the truncated model spaces.
The spectra obtained by using the $\alpha+{}^2n+{}^2n$ components, 
those using the $\alpha+{}^2n+{}^2n$
and the SM components, those of the full GCM calculation using the 
$\alpha+{}^2n+{}^2n$, SM, and $\mathrm{^6He}+{}^2n$ configurations are shown in the first, second, and third columns from the left.
The result obtained with the $\theta$-fixed GCM calculation 
using $\alpha+{}^2n+{}^2n$ at $\theta=\pi/2$, SM, and  
 $\mathrm{^6He}+{}^2n$ configurations are shown in the right column. 
  }
  \label{fig:Energy8he_sub}
\end{figure}

Also in $\Be$ system, the $2\alpha+{}^2n$ cluster component 
is dominant as 89\% and 92\% in $\Be(0_1^+)$ and $\Be(0_2^+)$,
respectively.
From these values, the dineutron cluster breaking component is estimated to be 11\% and 8\% in 
$\Be(0_1^+)$ and $\Be(0_2^+)$, respectively. 
For the ${}^6\mathrm{He}+\alpha$ cluster component, 
$\Be(0_2^+)$ has an approximately 60\% overlap with the $^6{\rm He}+\alpha$ cluster components, which indicates the spatial correlation between the dineutron cluster and an $\alpha$ cluster.

\subsection{Effects of dineutron breaking in $\He$}
\begin{table}[b]
  \centering
  \caption{R.m.s. radii 
  of the $0_1^+$ and $0_2^+$ states in $\He$, and the $0_1^+-0_2^+$ transition matrix elements $M$(IS0) obtained by the restricted GCM and the full GCM calculations}
  \label{tab:rad8he.sub}
  \begin{ruledtabular}
  \begin{tabular}{ccccccc}
                   & $r_m(0_1^+)$ & $r_m(0_2^+)$ &$M$(IS0)&\\
                   &    (fm)      &    (fm)      & (\si{fm^2})  &  &\\
                          \hline 
     $\alpha+{}^2n+{}^2n$     &    2.70      &    3.97      &  23.5  \\
     ($\alpha+{}^2n+{}^2n$) + SM  & 2.40      &    3.86      &  16.3  \\
         full                      &    2.40      &    3.91      &  15.5  \\
  \end{tabular}
  \end{ruledtabular}
\end{table}

In order to discuss the effects 
of the dineutron breaking components in $\He$, 
we perform restricted GCM calculations within truncated model spaces.
We calculate $\He$ wave functions by superposing only $\alpha+{}^2n+{}^2n$
configurations ($\alpha+{}^2n+{}^2n$ calculation) but not ${}^6\rm{He}+{}^2n$ and SM configurations. 
We also perform calculations by using $\alpha+{}^2n+{}^2n$
and SM configurations ($(\alpha+{}^2n+{}^2n)$+SM calculation) but not $\mathrm{^6He}+{}^2n$ configurations.

The energy spectra obtained with the $\alpha+{}^2n+{}^2n$
calculation and the ($\alpha+{}^2n+{}^2n$)+SM
calculation are compared with those of the full calculation  
in Figure~\ref{fig:Energy8he_sub} (three columns from the left).
Comparing the results of the $\alpha+{}^2n+{}^2n$ and
($\alpha+{}^2n+{}^2n$)+SM calculations, one can see that
the mixing of the SM configuration into the $\alpha+{}^2n+{}^2n$ configurations
causes a significant energy gain as \SI{1.7}{MeV} of the ground state, 
which is contributed by the spin-orbit interaction. 
This means that the dineutron breaking due to the spin-orbit interaction 
plays an essential role 
in the binding of the $\He$ system. 
It also contributes to a \SI{0.78}{MeV} energy gain in the $0_2^+$ state, 
but it is not as large as the ground state
because the effect of the spin-obit attraction is weaker 
for spatially developed dineutron clusters in $\He(0_2^+)$.

\begin{figure*}[t]
  \centering
  \includegraphics[width=\linewidth]{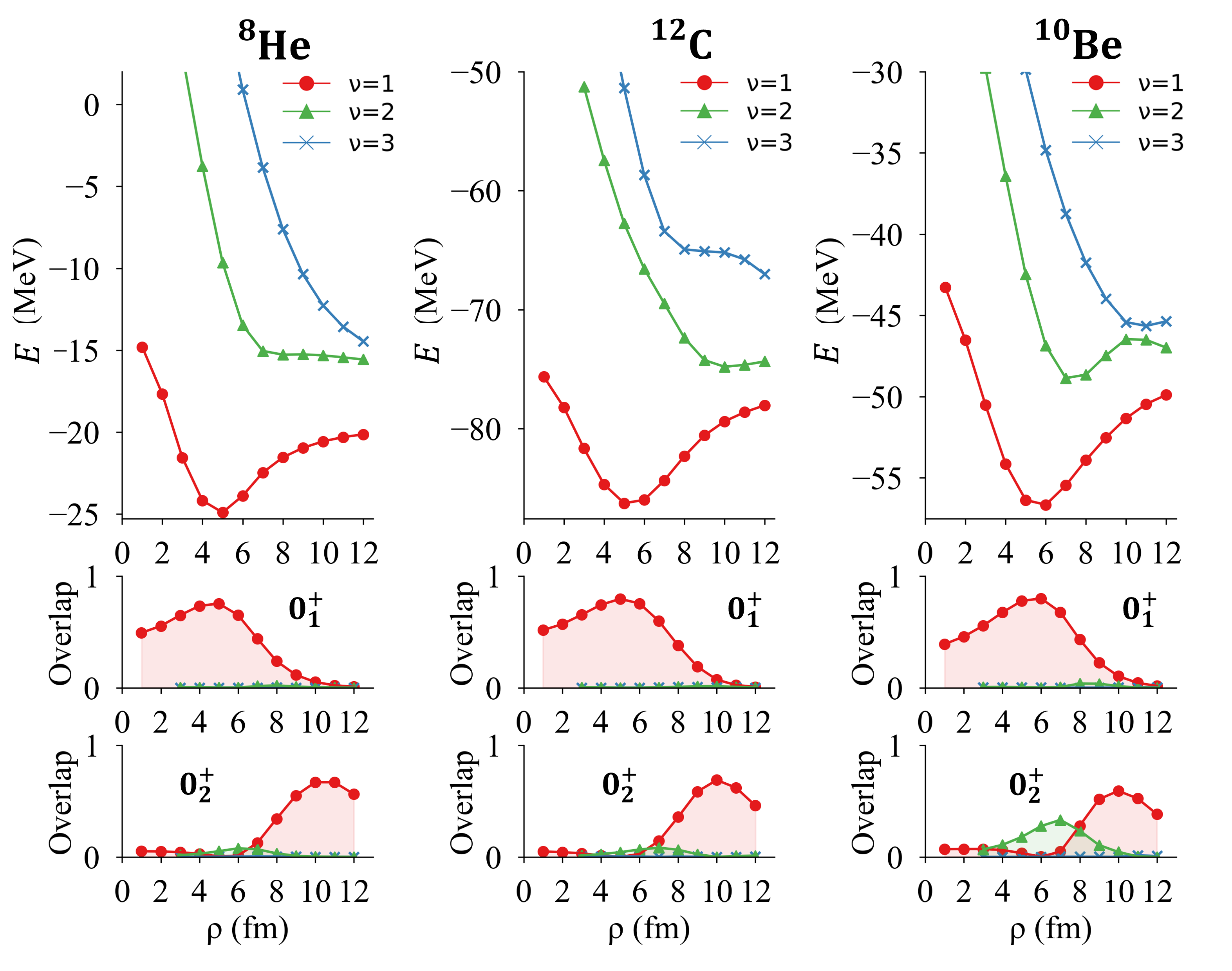}
  \caption{(Top) Energy spectra of $\He, \C$ and $\Be$ 
of the $\rho$-fixed 3-body states plotted as functions of $\rho=\rho_0$. 
 (Middle) Squared overlaps of the 
$\rho$-fixed states
$\ket{\varphi_{\nu}(\rho_0)}$ with the $0_1^+$ states.
 (Bottom) Those with the $0_2^+$ states.
The results for $\nu=1$, $\nu=2$, and $\nu=3$ are shown by 
red circles, blue triangles, and green cross symbols, respectively.}
  \label{fig:adiabatic.Eovlp}
\end{figure*}

In comparison of the ($\alpha+{}^2n+{}^2n$)+SM result with 
the full calculation, one can assess the effects of 
the $\mathrm{^6He}+{}^2n$ cluster component, 
in which two neutrons form a spatially developed dineutron
but the other two neutrons stay in the $p_{3/2}$ orbit around the 
$\alpha$. 
The mixing of this $\mathrm{^6He}+{}^2n$ component generates
an additional energy gain as $\SI{0.78}{MeV}$ 
for the $0_1^+$ and $0_2^+$ states.

The results of $r_m$ and $M$(IS0) obtained by the
$\alpha+{}^2n+{}^2n$ calculation and those by the ($\alpha+{}^2n+{}^2n$)+SM calculation are listed in Table \ref{tab:rad8he.sub} together with those of the full calculation.
Comparing the $\alpha+{}^2n+{}^2n$ results with the full
calculation, it is found that the dineutron cluster breaking gives 
a significant contribution to the system size shrinking, in particular,
of the ground state; an approximately 10\% reduction in $r_m$, and a 30\% reduction in $M$(IS0).
These results indicate again that the spin-orbit interaction 
causes the deeper binding involving the dineutron cluster breaking
as previously shown in the energy spectra in Figure~\ref{fig:Energy8he_sub}.

Although the dineutron cluster breaking gives significant effects on the energies and nuclear size, 
but qualitative properties of $\He(0^+_1)$ and $\He(0^+_2)$ are similar between calculations
with and without the dineutron cluster breaking.
In other words, the $\alpha+{}^2n+{}^2n$ model can describe leading properties of $\He$, but 
combining it with the dineutron cluster breaking configurations is essential
for a detailed description of 
$\He(0^+_1)$ and $\He(0^+_2)$ structures.

\subsection{Radial behavior of 3-body cluster structures}
In order to clarify the monopole excitation modes in $\He$, $\Be$, and $\C$, we 
prepare $\rho$-fixed 3-body states and take overlaps with 
the $0^+_1$ and $0^+_2$ states obtained 
by the full GCM calculations.  
The $\rho$-fixed 3-body states are constructed by the 
GCM calculation 
with the $\rho=\rho_0$ constraint as 
\begin{equation}
  \begin{split}
    &\ket{\varphi_{\nu}(\rho_0)} \\
    =& \int d\gamma \, d\theta \, f_{\nu}(\gamma, \theta) 
   \hat P^{0+} \ket{C_1,C_2,C_3; \rho = \rho_0, \gamma, \theta},
  \end{split}
\end{equation}
where the label $C_1,C_2,C_3$ are $\alpha+{}^2n+{}^2n$, 
$3\alpha$, and $2\alpha+{}^2n$ for $\He$, $\C$, and $\Be$, respectively, 
and $\nu = 1,2$ is the label for the $\nu$th state in the $\rho$-fixed subspace.
For each $\rho_0$ value, 
coefficients $f_{\nu}(\gamma, \theta)$ are determined by solving 
the Hill-Wheeler equation for the generator coordinates ($\gamma$ and $\theta$), and energy spectra $E_\nu(\rho_0)$ are obtained.
We also calculate the overlaps of these $\rho$-fixed states
$\ket{\varphi_{\nu}(\rho_0)}$ with the $0^+_1$ and $0^+_2$ states
obtained by the full GCM calculations.

Figure~\ref{fig:adiabatic.Eovlp} shows the $\rho$-fixed results of the energy spectra
$E_\nu(\rho_0)$ for $\nu=1,2,3$ and 
the squared overlaps of $\varphi_{\nu}(\rho_0)$ with the full GCM wave functions
plotted as 
functions of $\rho_0$.
In all of $\He$, $\C$, and $\Be$, the ground states 
are almost exhausted by the $\nu=1$ states; 
they have the maximum overlap with the lowest state ($\nu=1$) at the energy minimum around $\rho=5$--$\SI{6}{fm}$ and distribute along the $\nu=1$ energy curve.

The $0_2^+$ states of $\He$ and $\C$ are approximately exhausted by the $\nu=1$ states and 
have large overlaps around $\rho=\SI{10}{fm}$.
This means that the $\He(0_2^+)$ and  $\C(0^+_2)$ are the radial excitation modes along $\rho$, and this is one of the similarities between $\He$ and $\C$.
In contrast, $\Be(0_2^+)$ is not simply described by the $\nu=1$ states but has
significant overlaps with $\nu=2$ states around $\rho=\SI{7}{fm}$.
This is different from the cases of $\He(0_2^+)$ and  $\C(0^+_2)$, and indicates an excitation from $\nu=1$ to  $\nu=2$ states, rather than the radial excitation.
The reason for this difference between $\He$, $\C$ and $\Be$ can be understood by the energy cost for the 
 $ \nu=1\to \nu=2$ excitation compared with the $\rho$ excitation along the $\nu=1$ states. 
The $\nu=2$ energy curve in $\Be$ shows an energy pocket around $\rho=\SI{7}{fm}$, and its minimum energy 
approximately degenerates with the $\nu=1$ energy at $\rho=\SI{12}{fm}$, 
meaning that the $\nu=2$ excitation occurs with the energy cost as small as the $\rho$ excitation 
and contributes to $\Be(0_2^+)$.
In the cases of $\He$ and $\C$, the $\nu=2$ energy curves do not show such a pocket 
and their energies are higher than the $\nu=1$ energy curve. 
As a result of the much energy costs for the $\nu=2$ states,  $\He(0_2^+)$
and $\C(0_2^+)$ have the $\nu=1$ dominance feature without the $\nu=2$ mixing.

\section{\label{sec:disc}Discussions}
\subsection{Detailed analyses of cluster structures in $\He$}
\subsubsection{Spatial configurations of the $\alpha+{}^2n+{}^2n$ clusters in $\He(0_2^+)$}
\label{sp.3B}
\begin{figure}[ht]
  \centering
  \includegraphics[width=\linewidth]{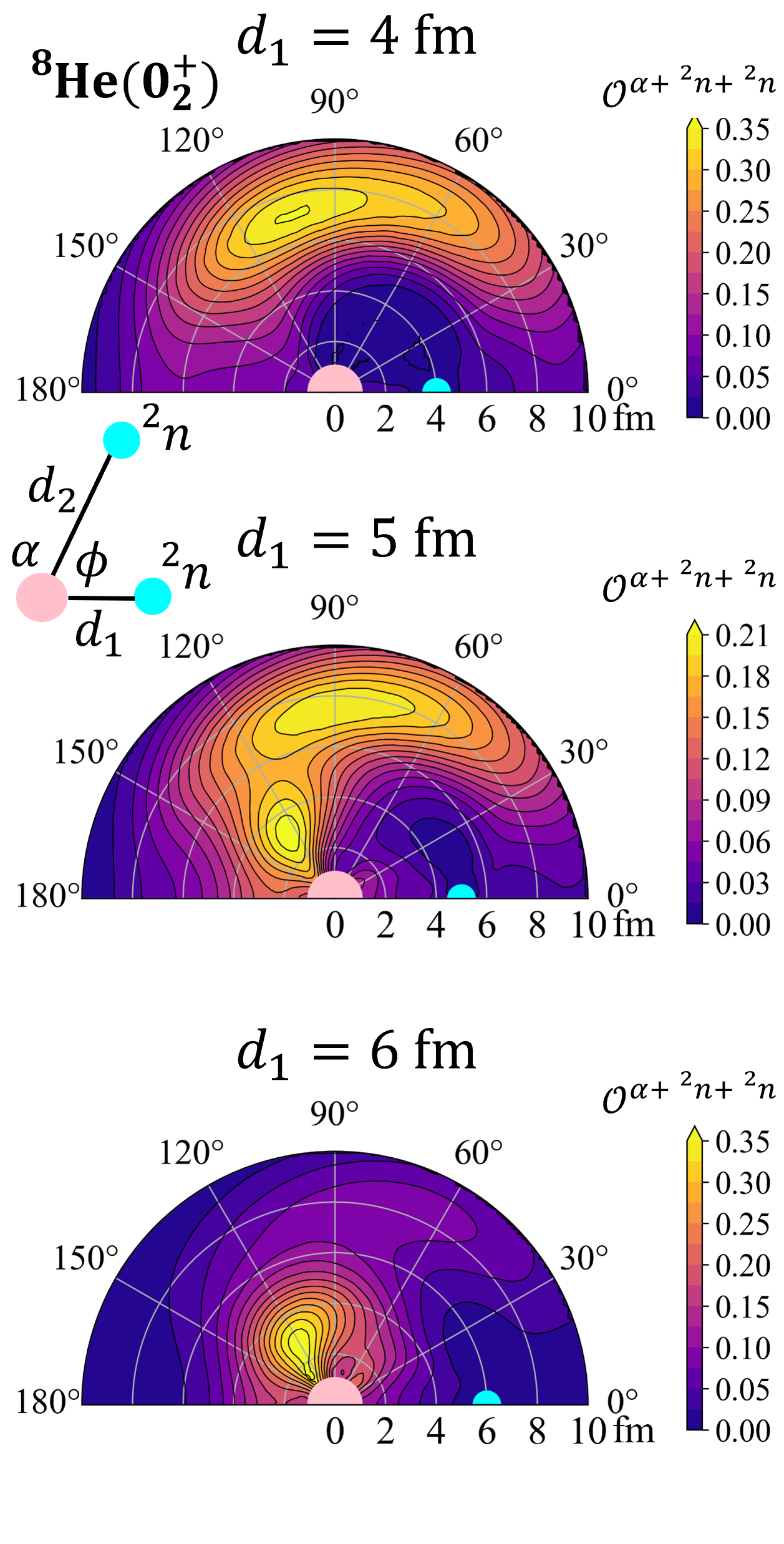}
  \caption{
  Spatial distribution of three clusters in $\He(0^+_2)$.
 The squared overlap ${\cal O}^{\alpha+{}^2n+{}^2n}(d_1,d_2,\phi)$ 
for fixed $d_1$ values at $d_1 = 4$~fm (top), 5~fm (middle),
and 6~fm (bottom) 
are plotted on the $(d_2,\phi)$ plane. 
The center positions of the $\alpha$ cluster 
and the ${}^2n$ cluster are displayed  
at the origin (pink circles) and at $\bm{d}_1=(d_1,\phi=0)$ (blue circles), respectively. 
  \label{fig:8heBO}}
\end{figure}

As mentioned in the $\rho$-fixed analysis in 
the previous section, $\He(0_2^+)$ is regarded as the 
radial excitation mode similar to $\C(0_2^+)$. 
We here give further discussions on the spatial configurations of the $\alpha+{}^2n+{}^2n$ cluster component
in $\He(0_2^+)$.
We calculate the squared overlap ${\cal O}^{\alpha+{}^2n+{}^2n}(\bm{d}_1,\bm{d}_2)$
of the $\alpha+{}^2n+{}^2n$ cluster configuration with $\He(0^+_2)$ obtained by the full GCM calculation as,
\begin{equation}
  {\cal O}^{\alpha+{}^2n+{}^2n}(\bm{d}_1,\bm{d}_2) \equiv \left| \bra{\alpha+{}^2n+{}^2n;\bm{d}_1,\bm{d}_2}\ket{\He(0_2^+)} \right|^2,
\end{equation}
where $\bm{d}_1=\bm{R}_1-\bm{R}_3$ and $\bm{d}_2=\bm{R}_2-\bm{R}_3$ are coordinates of the two ${}^2n$ clusters measured from the $\alpha$ cluster.
3-body configurations in the $0^+$-projected states are specified by 
$d_1=|\bm{d}_1|$, $d_2=|\bm{d}_2|$, and the opening angle $\phi$ between $\bm{d}_1$ and $\bm{d}_2$, 
and therefore, we rewrite the overlap with the parameters 
 $(d_1, d_2, \phi)$ instead of $(\bm{d}_1,\bm{d}_2)$ as  ${\cal O}^{\alpha+{}^2n+{}^2n}(d_1,d_2,\phi)
={\cal O}^{\alpha+{}^2n+{}^2n}(\bm{d}_1,\bm{d}_2)$.
The results of the squared overlaps
are shown in Figure~\ref{fig:8heBO}.
The calculated values of  ${\cal O}^{\alpha+{}^2n+{}^2n}(d_1,d_2,\phi)$ 
for fixed $d_1$ values at $d_1 = 4,5$ and $\SI{6}{fm}$ are plotted on the $(d_2,\phi)$ plane in the top, middle, and bottom panels, respectively.
These plots show spatial distributions of 
one dineutron when the center positions of
the $\alpha$ cluster and the other dineutron are located at the origin
(pink circles)  and  $\bm{d}_1=(d_1,\phi=0)$ (lightblue circles), respectively.

\begin{figure}[!t]
  \centering
  \includegraphics[width=\linewidth]{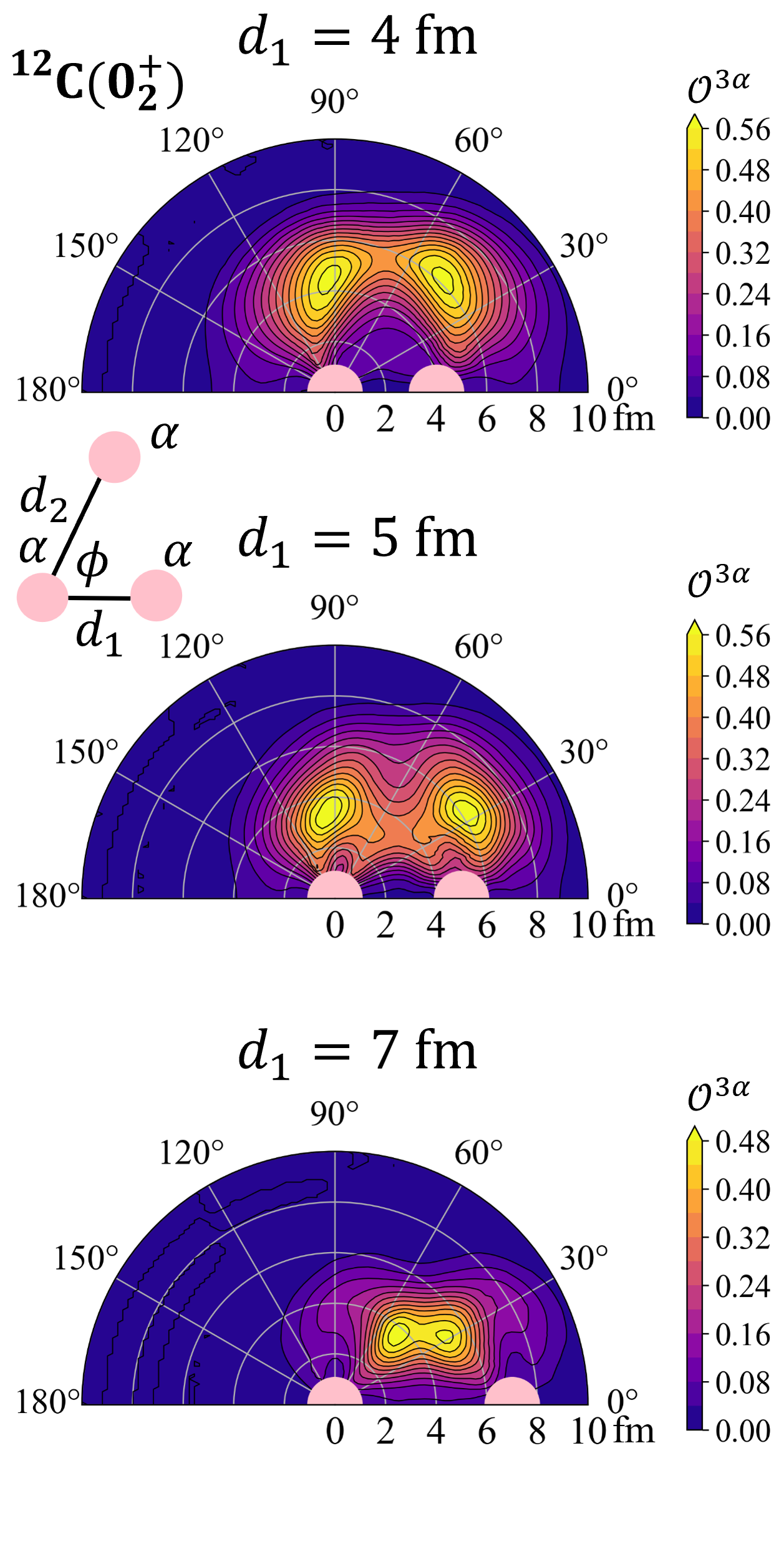}
  \caption{Same as Figure~3 but for $3\alpha$ clusters of
  in $\C(0_2^+)$. ${\cal O}^{3\alpha}(d_1,d_2,\phi)$ for $d_1=4,5,7 \, \si{fm}$
  are shown in the top, middle, and bottom panels, respectively. 
Center positions of the fixed two $\alpha$ clusters are shown by pink circles.}
  \label{fig:12CBO}
\end{figure}

In the $d_1 = \SI{4}{fm}$ case, the dineutron widely distributes along the angular direction around the $d_2=\SI{7}{fm}$ region, which shows $S$-wave behavior of the two dineutrons around the $\alpha$ cluster.
This $S$-wave feature is consistent with the results of the ${}^2n$-condensation model~\cite{kobayashi2013}.
In the $d_1=\SI{5}{fm}$ case, the $S$-wave $\alpha+{}^2n+{}^2n$ behavior still remains, but another peak appears around $d_2=\SI{3}{fm}$, $\phi=120\textrm{--}\ang{130}$.
This new peak exhibits 2-body-like $(\alpha+{}^2n)+{}^2n$ cluster structure, where one of two dineutrons  approaches
the $\alpha$ cluster and the other dineutron moves far from the $(\alpha+{}^2n)$ cluster, which consists of the correlated $\alpha$ and ${}^2n$ clusters.
Note that this $\alpha+{}^2n$ correlation is induced by the attraction of the spin-orbit interaction via the ${}^2n$ cluster breaking.
In $d_1 = \SI{6}{fm}$, there are no longer $S$-wave $\alpha+{}^2n+{}^2n$ structures but
the 2-body-like $(\alpha+{}^2n)+{}^2n$ cluster structure is dominant.

From this analysis, $\He(0_2^+)$ is characterized by 
two kinds of structures. One is the gas-like $\alpha+{}^2n+{}^2n$ structure of 
$S$-wave dineutrons, and the other is the 2-body-like  
$(\alpha+{}^2n)+{}^2n$ structure. 

\subsubsection{$\theta$-fixed analysis}
In order to discuss the role of the angle $\theta$ degree of freedom (DOF), 
we perform the $\theta$-fixed GCM calculation 
by fixing $\theta=\pi/2$ of $\alpha+{}^2n+{}^2n$ cluster configurations, 
which corresponds to 
isosceles triangular configurations with the $\alpha$ cluster at the vertex, as 
\begin{equation}
 \begin{split}
  & \ket{\theta \text{-} \mathrm{fixed} \,\, ^8\mathrm{He}(0_{\nu}^+)} \\
 =& \int d\rho \, d\gamma \, f_{\nu}^{3B}(\rho,\gamma) 
 \hat{P}^{0^+}\ket{\alpha+ {}^2n+ {}^2n;\rho,\gamma,\theta=\pi/2}  \\
 &+ f_{\nu}^{SM} \ket{\Psi^{SM}}  
 + \int da \, f_{\nu}^{2B}(a) \hat{P}^{0+}\ket{^6\mathrm{He}+{}^2n;a},
 \end{split} 
\end{equation}
where the coefficients $f_{\nu}^{3B}(\rho,\gamma), f_{\nu}^{SM}, f_{\nu}^{2B}(a)$ are determined as explained in Sec.~\ref{form.GCM}.
The energy spectra of the $\theta$-fixed calculation are shown 
in the right column of Figure~\ref{fig:Energy8he_sub}.
Comparing the energy spectra obtained by the $\theta=\pi/2$
fixing calculation with those of the full GCM calculation
without $\theta$ fixing, 
one can see a significant energy gain, in particular, 
for the $0^+_2$ state. This result indicates that the angular DOF is 
important for the $\alpha+{}^2n+{}^2n$ cluster structure in the $0_2^+$ state
and the superpositions of the $\alpha+{}^2n+{}^2n$ configurations 
along the angle $\theta$ is 
essential to describe the $S$-wave feature of the dineutron motion
around the $\alpha$ in $\He(0^+_2)$.

\subsubsection{Spatial development of the $\mathrm{^6He}+{}^2n$ cluster structure}
\begin{figure}[b]
  \centering
  \includegraphics[width=\linewidth]{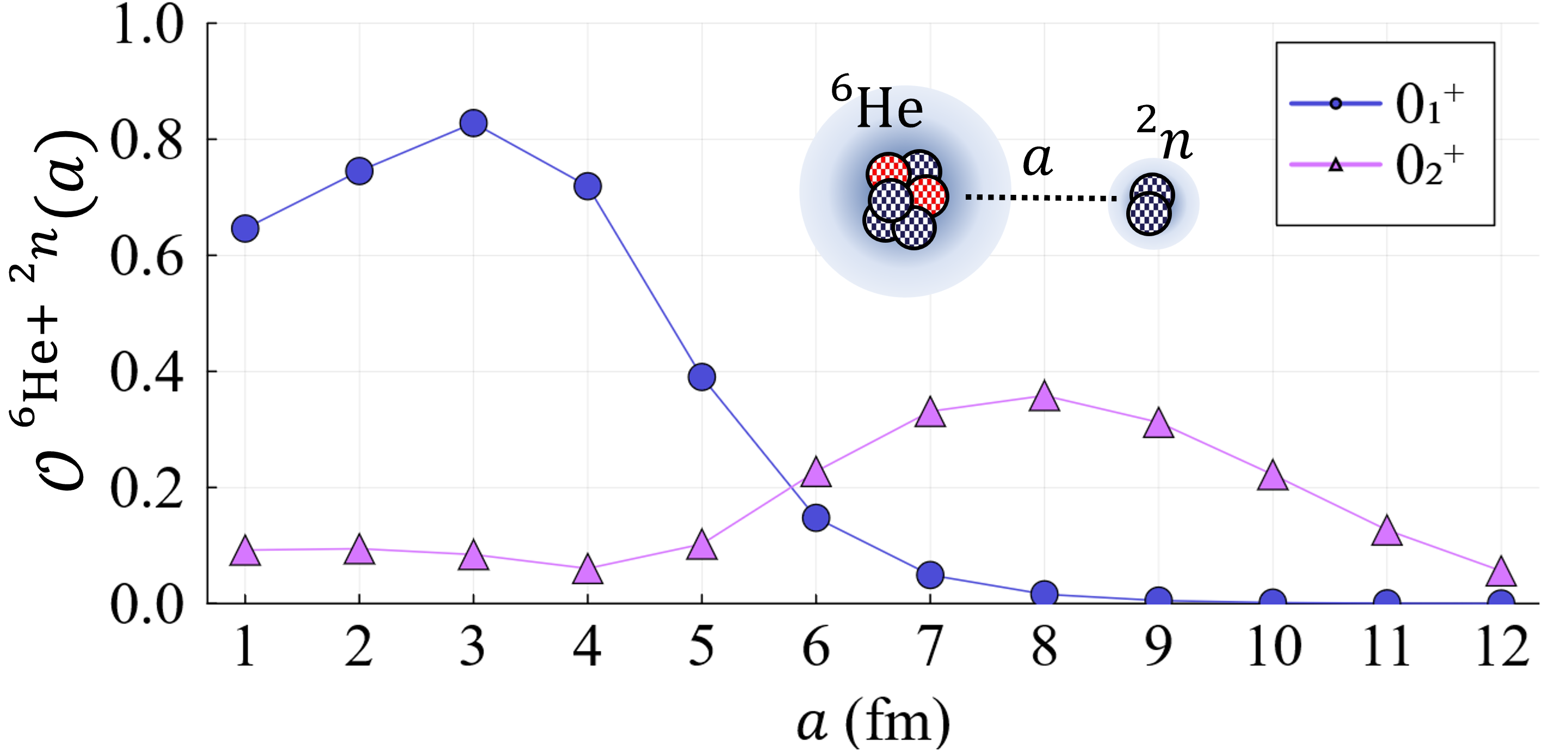}
  \caption{
  Squared overlaps ${\cal O}^{^6\textrm{He}+{}^2n}(a)$
  with each $^6\mathrm{He}+{}^2n$ configuration at distance $a$~fm
  for $\He(0_1^+)$ and $\He(0_2^+)$. 
  The results for $\He(0_1^+)$ and $\He(0_2^+)$ are plotted by blue circles and purple triangles, respectively. 
}
  \label{fig:2body}
\end{figure}
As discussed previously in the 3-body analysis (Sec. \ref{sp.3B}), 
$\He(0_2^+)$ contains the 2-body-like $(\alpha+{}^2n)+{}^2n$ component, which is induced by the attraction of the spin-orbit interaction via the ${}^2n$ cluster breaking.
To discuss the spatial distribution of the ${}^2n$ cluster around 
the $\mathrm{^6He}$-cluster, we calculate the squared overlap 
${\cal O}^{^6\textrm{He}+{}^2n}(a)$
of the $^6\mathrm{He}+{}^2n$ cluster configurations 
at the distance $a$ with the $\He(0^+_1)$
and $\He(0^+_2)$ wave functions obtained by the full GCM calculation as,
\begin{equation}
  \begin{split}
    {\cal O}^{^6\textrm{He}+{}^2n}(a) \equiv 
    \left| \bra{{}^6\mathrm{He}+{}^2n;a} \ket{\He(0_{1,2}^+)} \right|^2.
  \end{split}
\end{equation}

The calculated values of ${\cal O}^{^6\textrm{He}+{}^2n}(a)$ for the $0_1^+$ and $0_2^+$ states are shown in Figure~\ref{fig:2body}.
The $0_1^+$ state has large overlaps in the small distance region with a peak around $a = \SI{3}{fm}$ and rapid damping 
in $a> \SI{3}{fm}$, indicating the compact ${}^6\rm{He}+{}^2n$ cluster structures in the $0_1^+$ state.
It should be noted that this compact ${}^6\rm{He}+{}^2n$ component has large overlaps with the SM and compact $\alpha+{}^2n+{}^2n$ cluster configurations.
For the $0_2^+$ state, ${\cal O}^{{}^6\mathrm{He}+{}^2n}(a)$ shows a peak around $a=\SI{8}{fm}$ and a long tail toward large $a$ regions.
This result shows a weakly binding feature of the ${}^6\rm{He}+{}^2n$ cluster structure in $\He(0_2^+)$, that is, 
a widely distributing dineutron around the ${}^6\rm{He}$-cluster.
This is consistent with the 2-body-like structures $(\alpha+{}^2n)+{}^2n$ observed in the $\alpha+{}^2n+{}^2n$ analysis in Sec.~\ref{sp.3B}.

\subsubsection{Spatial configurations of $3\alpha$ clusters in $\C(0^+_2)$}
We analyze the $3\alpha$ cluster structure in $\C$
and discuss analogies and differences of 3-body 
cluster structures between $\C(0_2^+)$ and $\He(0_2^+)$. 
As done for $\He$,
we calculate the squared overlap of the $3\alpha$ cluster configurations ${\cal O}^{3\alpha}(d_1,d_2,\phi)$ as
\begin{equation}
{\cal O}^{3\alpha}(d_1,d_2,\phi)=
  {\cal O}^{3\alpha}(\bm{d}_1,\bm{d}_2) \equiv \left| \bra{\C(0_2^+)}\ket{3\alpha;\bm{d}_1,\bm{d}_2} \right|^2
\end{equation}
with $\bm{d}_1=\bm{R}_1-\bm{R}_3$, $\bm{d}_2=\bm{R}_2-\bm{R}_3$, 
$d_1=|\bm{d}_1|$, $d_2=|\bm{d}_2|$ and $\phi$ is the opening angle between $\bm{d}_1$ and $\bm{d}_2$.
The calculated values of ${\cal O}^{3\alpha}(d_1,d_2,\phi)$ for $d_1=4,5,$ and $\SI{7}{fm}$ are shown in the top, middle, and bottom panels of Fig.~\ref{fig:12CBO}, respectively.
In the result of $d_1 = \SI{4}{fm}$, the $\alpha$ cluster widely distributes along $\phi$ direction, which suggests the $S$-wave nature of the $\alpha$
cluster motion in $\C(0^+_2)$. This  $S$-wave behavior is one of the similarities with the $\He(0_2^+)$ structure.
However, the spatial extent is weaker in $\C(0_2^+)$ than in $\He(0_2^+)$ as seen in the $d_2$  distribution in ${\cal O}^{3\alpha}(d_1=4~\textrm{fm},d_2,\phi)$ compared with that in ${\cal O}^{\alpha+^2n+^2n}(d_1=4~\textrm{fm} ,d_2,\phi)$; 
the former and the latter shows significant amplitudes in the $d_2=4\textrm{--}6$~fm regions and $d_2=6$--$10$~fm regions, respectively. 
In other words, the weakly binding 3-body feature of the $\alpha+{}^2n+{}^2n$ cluster structure in $\He(0^+_2)$ is enhanced 
compared with that of the $3\alpha$ cluster structure in $\C(0^+_2)$. This feature can also be seen in the larger $r_m$ of $\He(0_2^+)$ than that of $\C(0_2^+)$.
Another difference from $\He(0^+_2)$ is that the 2-body-like $(\alpha+\alpha)+\alpha$ structure does not appear in $\C(0_2^+)$. 
As seen in the result of $d_1 = \SI{7}{fm}$, the amplitude concentrates at $d_2 \sim \SI{4}{fm}$ and $\phi \sim \ang{30}$,
which corresponds to a collapsed isosceles triangle configuration of $3\alpha$ clusters instead of the 2-body-like  structure. 
This feature can be understood by the difference in the inter-cluster interactions between the $\alpha+\alpha$ and $\alpha+^2n$ systems. Namely,
the $\alpha+\alpha$ correlation is unfavored because of the nucleon Pauli blocking effect, and the absence of the spin-orbit attraction at short distances.

\section{\label{sec:sum}summary}
  We investigated the structures of the $0_1^+$ and $0_2^+$ states of $\He$ with a microscopic cluster model combined with the cluster breaking configurations, and analyzed the $\alpha+{}^2n+{}^2n$  cluster
structures and the dineutron breaking effects 
induced by the spin-orbit interaction.
For describing dineutron motions in detail, 
we superposed the $\alpha+{}^2n+{}^2n$ cluster 
and ${}^6\rm{He}+{}^2n$ cluster configurations in the GCM framework, 
and also incorporated the mixing of the shell-model (SM) configuration. 

We calculated the energies and r.m.s. radii of the $0_{1,2}^+$ states and the matrix element of the $0_1^+ - 0_2^+$ transition $M$(IS0), and found that 
the calculated results are in reasonable agreement with the experimental values and other theoretical calculations.
In the results, we obtained a compact structure of the $0_1^+$ state, and the developed $\alpha+{}^2n+{}^2n$ cluster structure 
in the $0_2^+$ state.

The $\alpha+{}^2n+{}^2n$ cluster component has dominant contributions in both $\He(0_1^+)$ and $\He(0_2^+)$.
The $0_1^+$ state is a mixture of the compact $\alpha+{}^2n+{}^2n$ cluster structures and the SM component, 
while the $0_2^+$ state has the 
spatially developed $\alpha+{}^2n+{}^2n$ structures and also 
the $^6\rm{He}+{}^2n$ structures, showing the spatial correlation between the $\alpha$ and ${}^2n$ cluster in the 3-body cluster dynamics.

In comparison of the results obtained 
with and without the ${}^2n$ cluster breaking, it was revealed that 
the dineutron cluster breaking gives important contributions 
to the energy gain and size shrinking of the $0_1^+$ and $0_2^+$ states.

We also calculated the $0_1^+$ and $0_2^+$ states of $\C$ and $\Be$ by respectively applying the $3\alpha$ and $2\alpha+{}^2n$ cluster models combined with cluster breaking configurations, and compared the properties of the monopole excitations with those of $\He$.
It was found that the excitation to 
the $0_2^+$ states in $\He$ and $\C$ are regarded as the radial excitation modes along the hyperradius $\rho$, whereas the $0_2^+$ state in $\Be$ 
is not described by the radial excitation but shows different characters
from those in $\He$ and $\C$.

\bibliographystyle{apsrev4-2.bst}
\bibliography{mainbib.bib}
\end{document}